# Direct experimental observation of periodic intensity modulation along straight hollow core optical waveguides

T. Pfeifer

M. C. Downer

*downer@physics.utexas.edu*

*Department of Physics, The University of Texas at Austin, Austin, Texas 78712*

We report the direct observation of periodic intensity modulation of a laser pulse propagating in a hollow core waveguide. A series of equally spaced plasma sparks along the gas-filled capillary is produced. This effect can be explained by the beating of different fiber modes, which are excited by controlling the size of the focal spot at the capillary entrance. As compared to an artificial modulated waveguide structure, our presented approach represents an easier and more flexible quasi-phase-matching scheme for nonlinear-optical frequency conversion.

*OCIS codes:* 060.4370, 270.1670, 320.7110, 270.4180

## 1. Introduction

Intensity-modulating laser light along its propagation path has important consequences for nonlinear optics. Very recently, progress on quasi-phase matching of high-harmonic generation (HHG) has been achieved by using periodic intensity variation of a laser pulse propagating along a modulated hollow-core waveguide.[1,2,3] There are many more examples in nonlinear optics where quasi-phase matching can be employed to yield higher nonlinear-optical conversion efficiencies.[4,5] In another field of research, the problem of dephasing in laser-wakefield acceleration[6,7] can also be addressed by employing a suitably modulated laser pulse intensity. It enables the accelerated electron bunches to 'rephase' with the plasma wake at those positions where it is less intense and to be in phase for acceleration at the positions where the wake field is strongest.

A technical solution to create a periodic intensity variation is the modulated waveguide,[1,2] which provides a fixed modulation period for the intensity of the laser pulse. For both applications mentioned above however, an adjustable periodicity of the modulation would be favorable in order to optimize the processes. Quasi-phase matching could then be achieved for a selectable soft-x-ray photon energy in the HHG case and efficient matching to the electron kinetic energy for laser-wakefield acceleration could be achieved.

In this paper, we present and experimentally demonstrate a way of modulating the intensity along a waveguide *without* employing a modulated waveguide. The basic idea is the excitation of more than one (hollow) fiber mode at the entrance of the waveguide. Since different fiber modes possess different longitudinal wavevectors, the superposition of two or more leads to a characteristic beating behavior of the intensity along the propagation axis. In essence, the modulated waveguide can be regarded as a fiber-mode filter, which can be replaced by direct selective initial (at the entrance of the fiber) excitation of the suitable fiber modes. As we will show, it is very easy to excite a superposition of the two lowest-order modes in order to produce an intensity modulation just by choosing the appropriate size of the focus at the entrance of the fiber. In the future, tailoring the spatial phase of the laser prior to focusing[8] allows a straightforward way to excite an arbitrary superposition of fiber modes, leading to an adjustable period of the on-axis intensity modulation. Our approach even allows its implementation in plasma waveguides,[9] which can be used to increase the acceleration length for electrons in laser-wakefield acceleration.

## 2. Mechanism

As has been calculated by Marcatili and Schmeltzer,[10] for $\lambda \ll a$ ($\lambda$: wavelength, $a$: inner radius of the hollow waveguide with refractive index of the cladding $\nu$) the hollow fiber modes in cylindrical coordinates can be approximately written as

$$\begin{aligned} E_{rnm} &= J_{n-1}(u_{nm}\frac{r}{a})\cos(n\theta)\exp\left(i(\gamma z - \omega t)\right), \\ E_{\theta nm} &= J_{n-1}(u_{nm}\frac{r}{a})\sin(n\theta)\exp\left(i(\gamma z - \omega t)\right), \\ E_{znm} &= 0 \end{aligned} \quad (1)$$

where

$$J_{n-1}(u_{nm}) = 0,$$
$$\gamma = k\left(1 - \tfrac{1}{2}\left(\tfrac{u_{nm}\lambda}{2\pi a}\right)^2\left(1 - \tfrac{i(\nu^2+1)}{2\pi a\sqrt{\nu^2-1}}\right)\right), \quad (2)$$

with $k = 2\pi/\lambda$ being the free-space propagation constant, $u_{nm}$ the $m$th root of the Bessel function of the first kind $J_{n-1}$, and $\gamma$ the complex propagation constant. Its imaginary part describes the attenuation of light propagating in a hollow fiber whereas its real part is the longitudinal wave-vector $k_z$. Different fiber modes thus have different longitudinal wave-vectors.

If we now consider a superposition of different modes, a beating along the propagation axis ($z$-axis) will occur. The transverse intensity distribution of the beam traveling along the capillary will continuously change. In particular, if we consider only two modes being excited, the transverse intensity pattern will reappear with a given periodicity $\Delta z$ along the propagation axis:

$$\Delta z = \frac{2\pi}{\Delta k_z}, \qquad (3)$$

with $\Delta k_z$ denoting the longitudinal wave-vector difference of the two fiber modes under consideration. The radial intensity distributions of the two lowest order modes $EH_{11}$ and $EH_{12}$ are plotted in Fig. 1a. Fig. 1b shows the radial intensity distributions $I(r) = |\mathbf{E}_{11}(r) \pm f\mathbf{E}_{12}(r)|^2$ for the additive (+) and subtractive (−) interference case, respectively. Although the mixing ratio of higher to lower order mode is only 1:25 (in terms of total power content) ($f=0.2$) for this example, the two distributions are strikingly different. In the case of additive interference the beam is confined to a small region very close to the axis, which results in an increased on-axis intensity. For the subtractive interference, which occurs at an axial position $\Delta z/2$ away, the beam power is more evenly distributed, decreasing the on-axis intensity. A ringlike intensity distribution is obtained instead. This example illustrates the possibility to create a periodic intensity distribution along the propagation axis without manipulation of the structural properties (diameter) of the waveguide.

How can we excite two fiber modes in an effective manner? As has been found by Abrams,[11] the optimal focus size for which coupling from a gaussian beam into the lowest order ($EH_{11}$) fiber mode is most efficient is given by

$$w = 0.64\ a, \qquad (4)$$

where $w$ the $1/e^2$ intensity radius of the beam. At this beam diameter, 98% of the incoming laser beam power is coupled into the $EH_{11}$ mode. If the beam size is chosen differently from that value, higher order modes are excited. For our case of hollow waveguides, where the refractive index of the fiber cladding is higher than that of the core, higher order modes generally suffer significantly higher losses.[10] This is the reason why in most experiments aiming at maximum overall laser throughput, this scenario is tried to be avoided. In Fig. 2 we plot the amount of power or pulse energy coupled into the two lowest-order radially symmetric fiber modes from an input gaussian beam with given $w$. A maximum of 98% at the value given in Eq. (4) is obtained for the $EH_{11}$ mode. Most efficient coupling into the $EH_{12}$ mode occurs at a relative waist size of $w/a = 0.26$, with roughly 20% efficiency. The finding that we excite the $EH_{12}$ mode effectively for smaller gaussian beams becomes clear by regarding the additive interference case shown in Fig. 1. Due to on-axis constructive and off-axis destructive interference, the intensity distribution closely resembles a gaussian with a $w/a$ of $\sim 0.5$ which is smaller than the optimal one needed for efficient coupling into $EH_{11}$.

## 3. Experimental results

For the experimental demonstration, the hollow fiber was located inside a vacuum chamber, which was back-filled with 1 bar argon gas. Alignment to the laser was done by means of a 3-axis translation stage, enabling translation of the position of the focus in three dimensions, without changing the angle between propagation direction and fiber axis. The fiber was aligned angularly with respect to the beam by tilting the fiber mount. An 0.3 m plano-convex lens was used to focus onto the entrance of the fiber. The f-number was measured to be 14, resulting in a focal radius $w = 12$ $\mu$m. With the fiber inner radius $a = 25.5$ $\mu$m, we obtain a ratio of $w/a = 0.5$. Our laser is a commercial Ti:sapphire system from Spectra Physics, delivering pulses of 1 mJ energy, 10 ns pulse duration, 800 nm central wavelength at a 1 kHz repetition rate. A charge-coupled device (CCD) camera behind an 800 nm band-block filter acquired a photographic image of the capillary, which is shown in Fig. 3. A series of light emission centers (sparks) periodically lines up along the capillary axis. Since the fundamental radiation of the laser was blocked, the emission is attributed to plasma recombination light. Plasma generation in argon at wavelengths of 800 nm and pulse durations of several ns proceeds via an initial multiphoton (seed) process aided by collisional (avalanche) ionization.[12] The process is thus nonlinearly dependent on intensity, acting as a local intensity probe. We can therefore state that we could directly observe a periodic modulation of intensity along the propagation direction of the laser light. This observation has only experimentally been reported for planar waveguides and low intensity levels so far, where the modulation was detected with a near field probe.[13] The separation of the plasma maxima (i.e. the beating period) in our experiments $\Delta z$ is $\sim$2.7 mm, which is in good agreement with Eq. (3) considering excitation of $EH_{11}$ and $EH_{12}$ for the given fiber radius.

## 4. Simulation and Discussion

To give further evidence, we performed a computer simulation of the propagation of light in a hollow gas-filled glass capillary similar to the one reported in[14] for different parameters. The fiber dimensions were reduced to a diameter of 3.4 $\mu$m and the laser pulse duration reduced to 10 fs in order to keep the computation time at an acceptable level while still being able to obtain a qualitative comparison. In our model, we numerically integrated Maxwell's equation with a discrete differences approach on a two-dimensional cartesian grid. The ionization rate was calculated[15] at each time step and the number of newly produced electrons per unit volume was added to the corresponding local free-electron density. The important ratio of initial beam waist to fiber radius was chosen to be $w/a = 44\%$ and thus close to the experimental parameters. The result for the free-electron density produced from the laser pulse is shown in Fig. 4.

Again we obtain a periodic modulation of plasma density along the capillary axis. The smaller beating period of ∼65 $\mu$m can be understood from the reduced dimensions of the fiber. The expected beating period obtained from Eq. 3 for the population of the two lowest order fiber modes for the simulated geometry is 57 $\mu$m which is in agreement with the numeric result, confirming the validity of the mechanism. In addition to the on-axis plasma modulation, an off-axis modulation is visible, which exhibits secondary maxima at axial positions half-way in between the on-axis maxima. This can be attributed to the ringlike intensity distribution for the case of subtractive fiber-mode interference (Fig. 1). The effect of plasma defocusing on the observed behavior has been studied by varying the intensity of the pulses in the simulation and the influence on the periodicity of the phenomenon was negligible.

We would now like to discuss the relevance of fiber mode interference for quasi-phase-matching schemes. For any kind of nonlinear frequency conversion application, the efficiency of the process is governed by the phase mismatch

$$\Delta k_{\rm pm} = \sum_n k_n, \tag{5}$$

where the $k_n$ represent the contributing wave-vectors in the process. For the quasi-phase-matching case, we have to realize a modulation of the coupling between the fundamental and converted frequencies which is periodic along the generation path. This way, the conversion process into the signal field will occur only when the interference between propagating signal and newly generated signal is constructive. The period length $\Delta z$ can be converted into an effective wave-vector $\Delta k_{\rm eff} = 2\pi/\Delta z$. Quasi-phase matching is achieved when

$$\Delta k_{\rm pm} = m\Delta k_{\rm eff} \tag{6}$$

with the order of quasi-phase matching $m$ being an integer number. Above we showed that fiber mode interference can directly create periodic intensity modulations which are characterized by an effective wavevector $\Delta k_{\rm eff} = \Delta k_z$ (see Eq. (3)). Therefore excitation of two (or more) fiber modes can be directly exploited for quasi-phase matching applications.

## 5. Conclusion

In conclusion we experimentally showed that excitation of more than one fiber mode leads to a periodic modulation of the radial intensity distribution of a laser beam propagating inside a hollow core fiber. We demonstrated this effect experimentally by creating a series of equally spaced laser induced plasma sparks along the fiber axis. This finding has important implications for quasi-phase-matching applications and schemes for laser-wakefield acceleration. We would finally like to point out that our considerations are not restricted to hollow capillaries but certainly apply to any kind of guided geometry.

**Acknowledgment**

Financial support from the National Science Foundation FOCUS Center grant PHY-0114336 and U. S. Department of Energy grant DEFG03-96-ER-40954 is gratefully acknowledged.

**List of Figure Captions**

**List of Figures**

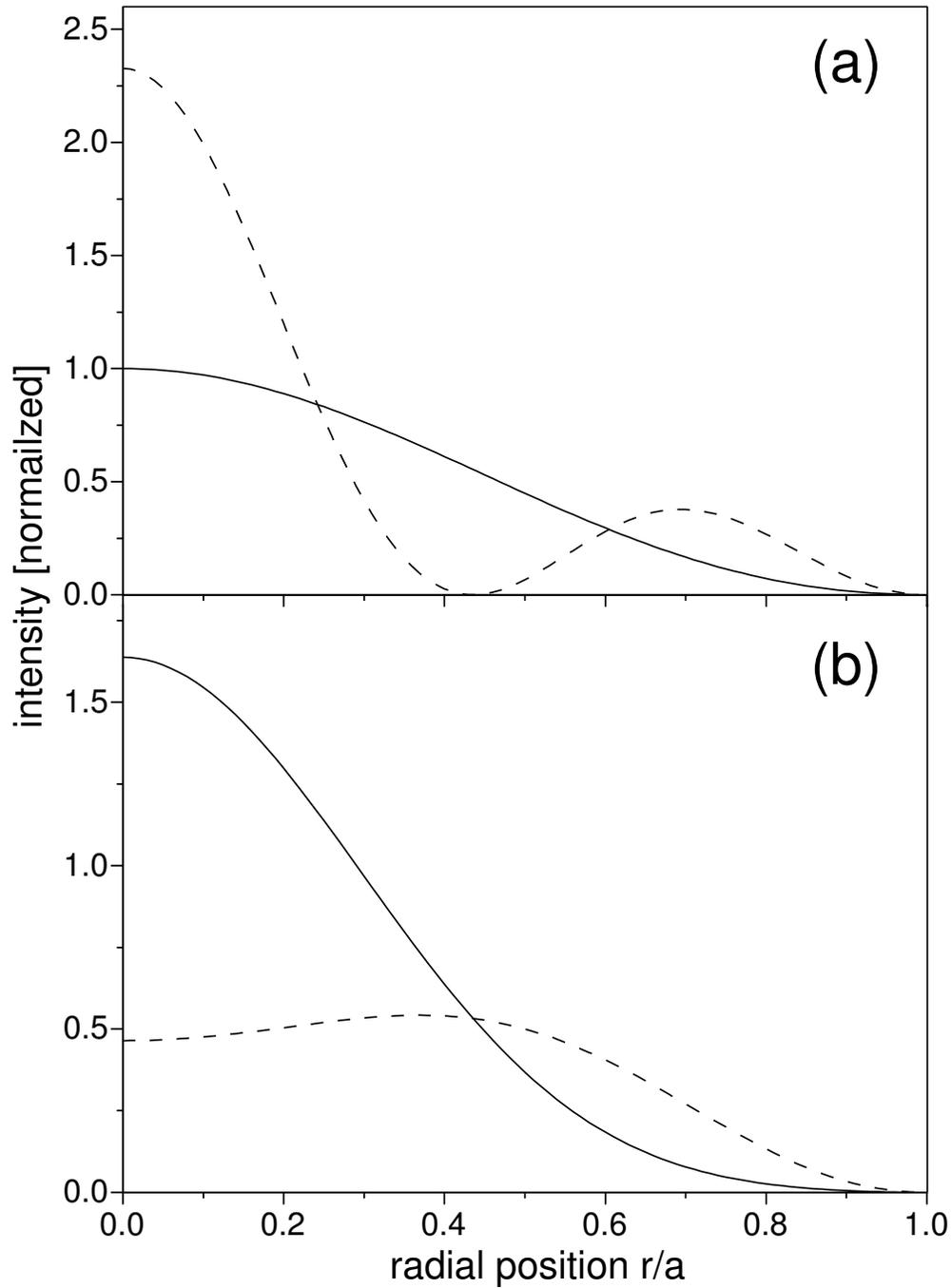

Fig. 1. Radial intensity distributions. (a): Pure hollow fiber modes $EH_{11}$ (solid line) and $EH_{12}$ (dashed line), normalized to integral power. (b): Superposition of both modes shown in (a) with mixing ratio 1:25 of $EH_{11}$ to $EH_{12}$ for additive (solid line) and subtractive interference (dashed line).

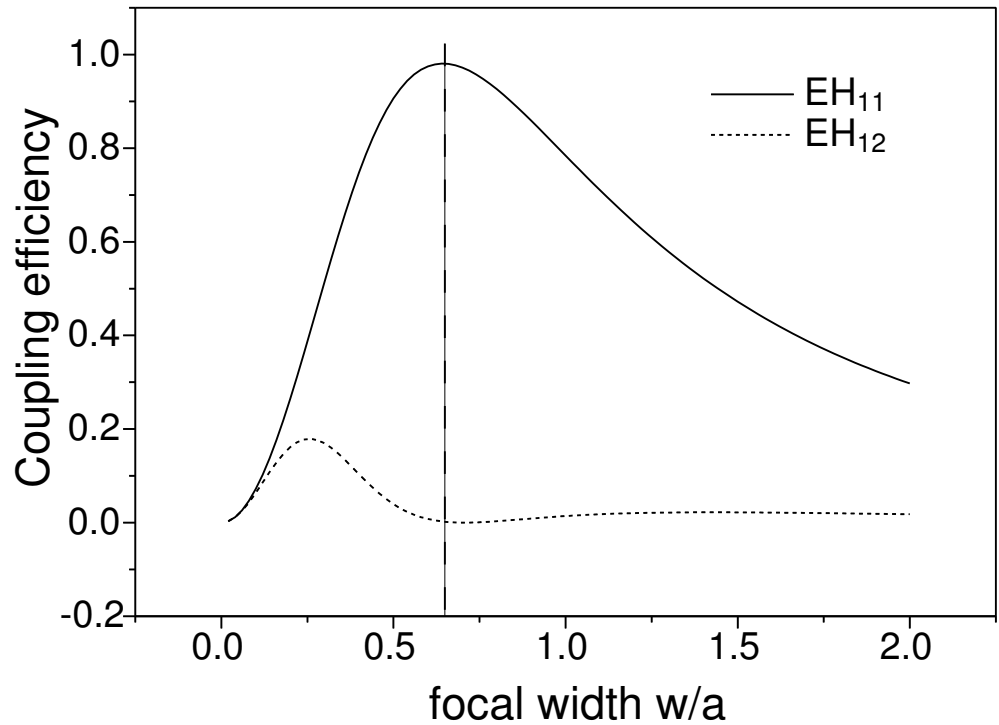

Fig. 2. Coupling a gaussian beam with width $w$ into hollow fiber modes (fiber inner radius $a$). For $EH_{11}$ (solid line) a maximum of 98% efficiency is obtained for $w/a = 0.64$. At smaller input beam diameters, higher order modes are excited, in particular $EH_{12}$ (dashed line).

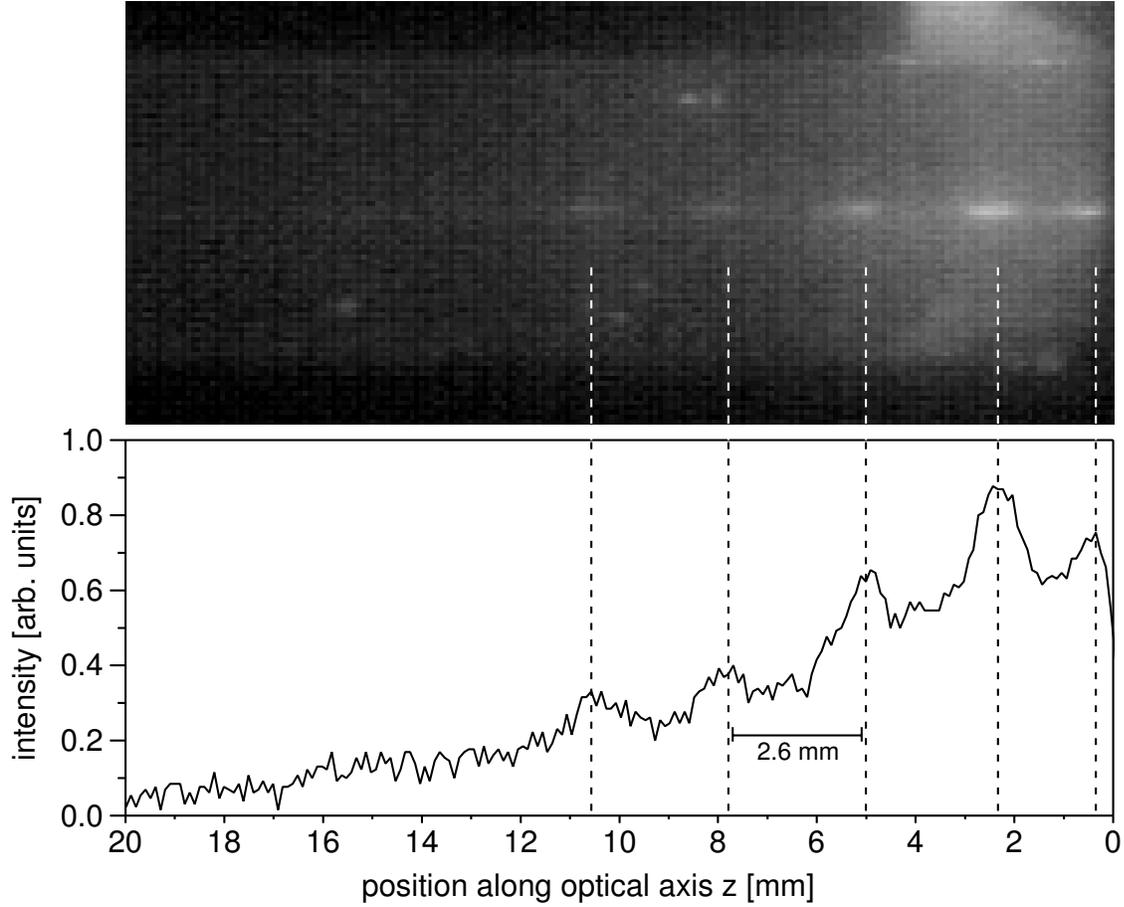

Fig. 3. Photographic image (above) and axial lineout (below) of the periodic plasma sparks created by fiber mode beating. The laser enters the capillary from the right-hand side. The distance between two on-axis plasma maxima is in agreement with the calculated value (see Eq. (3)) of 2.6 mm (indicated in figure) for the contribution of the two lowest order relevant fiber modes $EH_{11}$ and $EH_{12}$. A periodic intensity modulation along the propagation path of the laser light can thus be achieved without a structured waveguide, which is important for quasi-phase matching applications in nonlinear optics. The positions of the intensity maxima are marked by dotted lines.

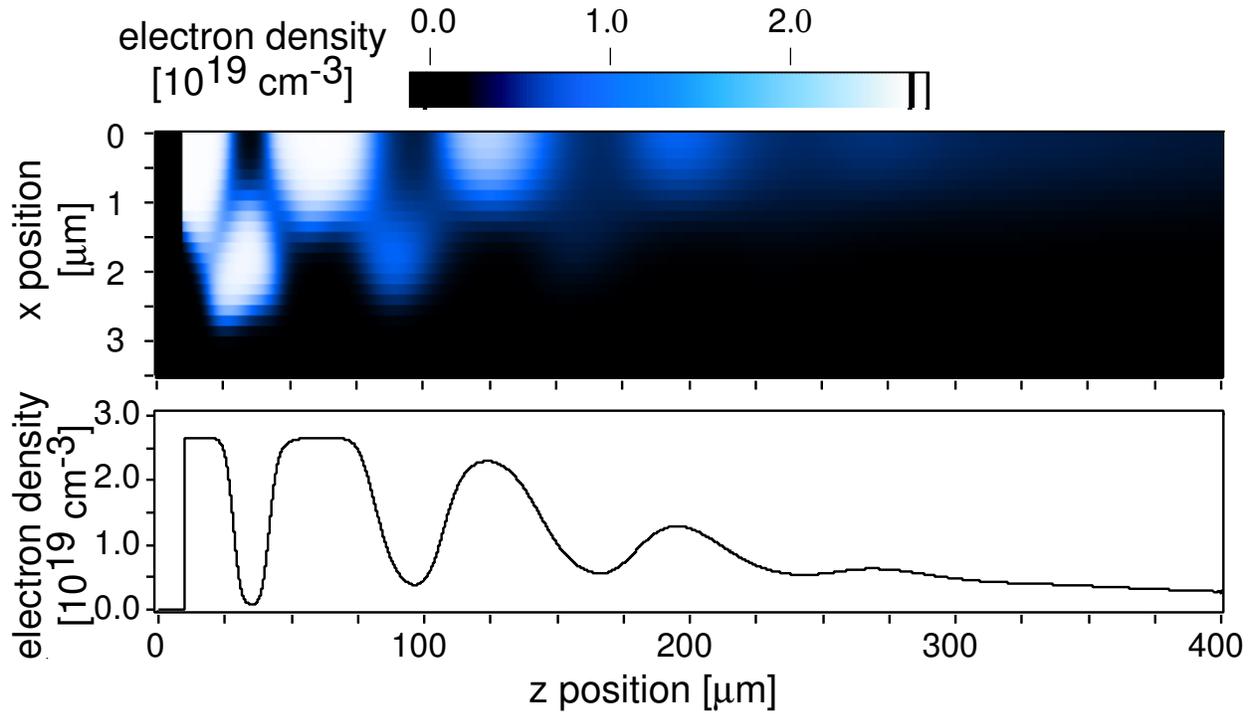

Fig. 4. Computer simulation results for light propagation in a gas filled capillary (pressure 1 bar). Shown is the free-electron density produced from the passage of the 10 fs laser pulse with an intensity of $10^{15}$ W/cm$^2$. As in the experiment, a periodic intensity modulation along the capillary axis leads to a periodic series of regions with high free electron (plasma) density. A less-pronounced off-axis intensity modulation occurs in anti-phase to the on-axis modulation. This is due to a transient ringlike intensity distribution when the on-axis electric fields of the fiber modes interfere destructively.